\documentclass[aps,amssymb]{revtex4}                  
\usepackage{graphicx}
\newcommand{\tr}{CMR }
\newcommand{\dfk}{TRFK }
\newcommand{\s}{blue }
\newcommand{\w}{red }
\newcommand{\g}{grey }
\newcommand{\q}{Q}

\begin{document}

\title{The Percolation Signature of the Spin Glass Transition
}

\author{J.~Machta}
\affiliation{
Physics Department,
University of Massachusetts,
Amherst, MA 010003 USA}
\email{machta@physics.umass.edu}
\author{C.M.~Newman}
\affiliation{
Courant Institute of Mathematical Sciences,
New York University,
New York, NY 100012 USA}
\email{newman@courant.nyu.edu}     
\author{D.L.~Stein}
\affiliation{
Physics Department and Courant Institute of Mathematical Sciences,
New York University,
New York, NY 100012 USA}
\email{daniel.stein@nyu.edu}

\begin{abstract}
Magnetic ordering at low temperature for Ising ferromagnets manifests
itself within the associated Fortuin-Kasteleyn (FK) random cluster
representation as the occurrence of a single positive density percolating
network. In this paper we investigate the percolation signature for Ising
spin glass ordering --- both in short-range (EA) and infinite-range (SK)
models --- within a two-replica FK representation and also within the
different Chayes-Machta-Redner two-replica graphical representation. Based
on numerical studies of the $\pm J$ EA model in three dimensions and on
rigorous results for the SK model, we conclude that the spin glass
transition corresponds to the appearance of {\it two\/} percolating
clusters of {\it unequal\/} densities.
\end{abstract}
\maketitle
\section{Introduction}
\label{sec:intro}

Ising type spin glass models, both of the short-range Edward-Anderson
(EA)~\cite{EA75} and the infinite-range Sherrington-Kirkpatrick
(SK)~\cite{SK75} varieties, have been studied for decades (for some recent
reviews, see~\cite{MPRRZ00} and~\cite{NSjpc03}). Nevertheless, to a large
extent, they remain a mystery --- especially the short-range variety, with
competing views as to the nature of their ordered phases at low
temperature~$T$~\cite{BY86,MPRRZ00,NSjpc03}. Indeed, from the perspective
of rigorous results, it is striking that there is no proof of broken
symmetry (e.g., of a nonzero EA order parameter) for any dimension~$d$ or
temperature~$T$.

Graphical representations such as the Fortuin-Kastelyn (FK) random cluster
model~\cite{KF69,FK72} are important tools in the study of spin systems.
They relate correlations in spin systems to geometrical properties of
associated random graphs.  Graphical representations are useful in
obtaining rigorous results concerning spin systems
(e.g.,~\cite{ACCN87,IN88}), they yield geometric insights into the nature
of phase transitions and they are the basis for powerful Monte Carlo
methods for simulating phase transitions~\cite{Swe,SwWa86,Wolff}. However,
graphical representations have, thus far, played a much less important role
in the study of spin glasses than they have for ferromagnets.

In this paper, we investigate two different graphical representations ---
the two-replica graphical representation of Chayes, Machta and
Redner~(CMR)~\cite{CMR98,CRM98} and a two-replica version of the
FK~representation (see Sec.~4.1 of~\cite{NS07}). Our purpose is to
understand the ``percolation signature'' of spin glass ordering within
these graphical representations. For ferromagnets, ordering corresponds to
the occurrence of percolating networks or clusters in the single replica
version of the FK representation. As we shall explain, we believe we have
elucidated the somewhat more complicated percolation signature for spin
glasses.

This should help in understanding better the differences between the nature of the 
phase transition in ferromagnets and in spin glasses. It is also our 
hope that for short-range models, this will be a significant
step towards developing a rigorous proof for spin glass ordering and eventually
lead to a clean analysis of the differences between short- and infinite-range
spin glass ordering. 

Understanding the percolation signature for spin glasses requires two
ingredients beyond what is needed for ferromagnets.  The first 
is the need to consider percolation within a two-replica
representation.  As mentioned, we consider two different such
representations --- one is the percolation of a certain class of bonds
(these are the ``blue bonds'' introduced and explained in
Subsection~\ref{tworepsection} below) in the CMR two-replica graphical
representation and the other is percolation of bonds that are doubly FK
occupied --- i.e., occupied in {\it both\/} replicas --- in the 
two-replica Fortuin-Kasteleyn (TRFK) representation. The two different
types of 
percolation, which we will often refer to simply as CMR and 
TRFK percolation, give relatively similar (but not identical) results,
with the major qualitative distinction occuring within the
SK spin glass.

The second ingredient, initially unexpected by us but in retrospect rather
natural, is that spin glass ordering corresponds to a more subtle
percolation phenomenon than simply the appearance of a percolating cluster
--- one that involves a pair of percolating clusters.  In the case of a
ferromagnet, there are general theorems~\cite{BK89} which ensure that when
percolation occurs, there is a unique percolating cluster, whether in
single or double replica representations.  It is also possible (by
averaging over disorder realizations) to show (see, e.g.,~\cite{GKN92}
and~\cite{NS07}) that the same conclusions are valid in the single replica
FK representation of spin glasses. However for blue bond percolation in the
(two-replica) CMR representation of spin glasses, both our numerical
evidence for the $d=3$ EA model and our rigorous results for the SK model
(in the CMR representation) show that at temperatures well above the spin
glass transition, there already is percolation, but that there are {\it
two\/} percolating networks which are equal in density (and presumably
otherwise macroscopically indistinguishable).  The SG transition
corresponds to the {\it breaking of indistiguishability\/} between the two
percolating networks --- in particular by having a nonzero difference in
densities. The latter feature also occurs for doubly occupied bonds in the
TRFK representation of the SK model, except that in that representation
there are no percolating networks at all above the transition temperature.

From a numerical perspective, spin glasses also pose major challenges.
Some of the numerical techniques, e.g., that of Swendsen and Wang
(SW)~\cite{SwWa86,WaSw88,WaSw05} based on graphical representations, such
as that of Fortuin and Kasteleyn, which have proven so useful for
ferromagnets~\cite{KF69,FK72}, are in principle applicable to spin
glasses. However, they are very inefficient in practice for values of $d$
and $T$ where ordering is believed to occur.  The CMR graphical
representation is related to a two-replica algorithm originally introduced
by Swendsen and Wang and developed by these and other
authors~\cite{ES88,Houdayer01,jorg05,jorg06}.  These authors have shown that
algorithms incorporating two-replica cluster moves are somewhat useful in
simulating spin glasses.  In particular, J\"org~\cite{jorg05,jorg06} has
shown that an algorithm based on a two-replica representation performs
reasonably efficiently for diluted spin glass models in three dimensions.
Two-replica cluster methods have also been successfully applied to Ising
systems in a staggered field~\cite{CRM98} and to the random field Ising model~\cite{MaNeCh00}.
The Monte Carlo method that we use takes advantage of the full
set of moves allowed by the CMR graphical representation.  These moves are
a superset of the moves used
in~\cite{SwWa86,WaSw88,Houdayer01,WaSw05,jorg05,jorg06}.

The paper is organized as follows.  In Sec.\ \ref{sec:grsg} we introduce
the idea of graphical representations, describe the CMR and TRFK
two-replica graphical representations and present properties of these
representations.  In Sec.\ \ref{sec:cg} we analyze both two-replica
representations on the complete graph --- i.e., for the SK spin glass.  In
Sec.\ \ref{sec:sim} we present numerical results for the three-dimensional
EA model.  The paper concludes with a discussion.

\section{Graphical Representations for Spin Glasses}
\label{sec:grsg}
\subsection{Fortuin-Kastelyn Graphical Representation}
\label{sec:gr}

Graphical representations for the Ising model originated with the work of
Fortuin and Kastelyn~\cite{KF69,FK72}. They were re-discovered and given a
physical interpretation by Coniglio and Klein~\cite{CK80}, applied as the
basis of a powerful algorithm for simulating the Ising model by Swendsen
and Wang~\cite{SwWa86,WaSw88,WaSw05} and then reformulated as a joint
spin-bond distribution by Edwards and Sokal~\cite{ES88}. Edwards and Sokal
introduced a joint distribution of spin variables $\{\sigma_x\}$ and bond
variables $\{\omega_{xy}\}$.  Here $\{x\}$ represents the set of sites
(vertices) of an arbitrary lattice (graph) and $\{xy\}$ the set of bonds
(edges).  The Ising spin variables take values $\pm 1$ and the bond
variables take values 0 or 1, or ``unoccupied'' and ``occupied'',
respectively.  The statistical weight ${\cal W}$ for the Edwards-Sokal
distribution is
\begin{equation}
\label{eq:es}
{\cal W}(\sigma,\omega;p)=p^{|\omega|}(1-p)^{N_b-|\omega|} \Delta(\sigma,\omega)
\end{equation}
Here $|\omega|=\sum_{\{xy\}} \omega_{xy}$ is the number of occupied bonds and $N_b$ is the total number of bonds on the lattice.  The factor $\Delta(\sigma,\omega)$ is defined by,
\begin{equation}
\label{eq:delta}
\Delta(\sigma,\omega)= \left\{ \begin{array}{l}
1 \mbox{   if  for every $xy$: }  \omega_{xy}=1\rightarrow
\sigma_x\sigma_y=1
 \\0 \mbox{ otherwise}\end{array} \right.
\end{equation}
The $\Delta$ factor requires that every occupied bond is satisfied.
Without the $\Delta$ factor we would have independent Bernoulli
percolation.  Given the choice, $p={\cal P}_{\rm FK}(\beta)=1-\exp(-2\beta
J)$, it is easy to verify that the spin and bond marginals of the
Edwards-Sokal distribution are the ferromagnetic Ising model
with coupling strength $J$ and the Fortuin-Kastelyn random cluster model,
respectively.

Bond and spin configurations in the ferromagnet contain
essentially the same information.  For example, the spin-spin correlation
function $\langle \sigma_x \sigma_y \rangle$ is equal to the probability
that sites $x$ and $y$ are connected by occupied bonds in the bond
representation,
\begin{equation}
\label{eq:corcon}
\langle \sigma_x \sigma_y \rangle = \mbox{Prob}\{x \mbox{ and } y \mbox{ connected}\}.
\end{equation}
This relationship implies that the phase transition in the spin system is
accompanied by a percolation transition in the bond system.

Given a typical equilibrium bond configuration one can construct a typical
equilibrium spin configuration by identifying connected components or
clusters and independently populating every spin in each cluster with one
randomly chosen spin type.  Similarly, given an equilibrium spin
configuration, an equilibrium bond configuration can be constructed by
occupying satisfied bonds with probability ${\cal P}_{\rm FK}(\beta)$.  The
equivalence between spin and bond configurations is the basis of the
Swendsen-Wang algorithm, which proceeds by successively creating spin
configurations from bond configurations and then bond configurations from
spin configurations.  It is easy to verify that this algorithm is ergodic
and satisfies detailed balance with respect to the Edwards-Sokal
distribution.  Power law decay of spin correlations at criticality imply
via Eq.\ (\ref{eq:corcon}) that the connected components of bond
configurations at criticality have a power law distribution of sizes.  The
efficiency of the Swendsen-Wang algorithm is due to the fact that the spin
system is modified on all length scales in a single step.

The FK representation is easily adapted to the $\pm J$ Ising spin glass.
(With minor modifications, it can also be adapted to Gaussian and other
distributions for the couplings, but we will generally not consider those
in this paper.)  The corresponding Edwards-Sokal weight is the same as
given in Eq.\ (\ref{eq:es}).  The $\Delta$ factor must still enforce the
rule that all occupied bonds are satisfied,
\begin{equation}
\label{eq:essg}
\Delta(\sigma,\omega;J)= \left\{ \begin{array}{l}
1 \mbox{   if  for every $xy$: }  \omega_{xy}=1\rightarrow
J_{xy}\sigma_x\sigma_y=1
 \\0 \mbox{ otherwise.}\end{array} \right.
\end{equation}
The spin marginal of the corresponding Edwards-Sokal distribution is the
Ising spin glass with couplings $\{J_{xy}\}$. Unfortunately, the
relationship between spin-spin correlations and bond connectivity is
complicated by the presence of antiferromagnetic bonds.  Specifically, one
has

\begin{eqnarray}
\langle \sigma_x \sigma_y \rangle=&\\
\mbox{Prob}\{x \mbox{ and} &y \mbox{ connected by even number of antiferromagnetic bonds}\}\nonumber\\
-\mbox{Prob}\{x \mbox{ and} &y \mbox{ connected by odd number of antiferromagnetic bonds}\}\nonumber.
\end{eqnarray}
It is no longer the case that the percolation of FK bonds implies long
range order~\cite{CoLiMoPe91}.  Two spins separated by a large distance may
usually be connected by occupied bonds but still be uncorrelated because
half the time the connection has an even number of antiferromagnetic bonds
and half the time an odd number of antiferromagnetic bonds. Indeed, FK
bonds percolate at a temperature that is well above the spin glass
transition temperature.  For the three-dimensional Ising spin glass on the
cubic lattice Fortuin-Kastelyn bonds percolate at $\beta_{{\rm
FK},p}\approx 0.26$~\cite{ArCoPe91} while the inverse critical temperature
is $\beta_c=0.89\pm 0.03$~\cite{KaKoYo06}.  Near the spin glass critical
temperature, the giant FK cluster includes most of the sites of the system.
For this reason, the Swendsen-Wang algorithm, though valid, is quite
inefficient for simulating spin glasses.

\subsection{The CMR Two-Replica Graphical Representation}
\label{tworepsection}

A conceptual difficulty of using the Fortuin-Kastelyn representation to
understand spin glass ordering is that FK clusters identify magnetization
correlations but the spin glass order parameter is not the magnetization.
Spin glass order is manifest in the Edwards-Anderson order parameter, which
can be defined with respect to two independent replicas of the system, each
with the same couplings $\{J_{xy}\}$.  The spins in the two replicas are
$\{\sigma_x\}$ and $\{\tau_x\}$, respectively, each taking values $\pm 1$.
The Edwards-Anderson order parameter, $q_{EA}$, is defined in
terms of the overlap,
\begin{equation}
Q \, = \, {N_s}^{\,-1}\sum_{\{x\}} \sigma_x \tau_x \ ,
\end{equation}
in the limit as the number of sites $N_s \to \infty$.
In general, $Q$ is a random variable whose maximum possible
value is $q_{EA}$, but in the case where (in the limit $N_s \to \infty$)
$\{\sigma_x\}$ and $\{\tau_x\}$ are drawn from a single pure state,
$Q$ takes on only the single value $q_{EA}$.
 
The two-replica graphical representation, introduced in~\cite{CMR98,CRM98},
explicitly relates spin glass order to geometry.  The associated
Edwards-Sokal joint distribution has, in addition to the spin variables,
$\{\sigma_x\}$ and $\{\tau_x\}$, two types of bond variables $\omega_{xy}$
and $\eta_{xy}$ each taking values in $\{0,1\}$.
 
The Edwards-Sokal weight is 
\begin{equation}
{\cal W}(\sigma,\tau,\omega,\eta;J)=B_{\rm \s}(\omega)B_{\rm \w}(\eta)
\Delta(\sigma,\tau,\omega;J)\Gamma(\sigma,\tau,\eta;J)
\end{equation}
where the $B$'s are Bernoulli factors for the two types of bonds,
\begin{eqnarray}
B_{\rm \s}(\omega)={\cal P}_{\rm \s}^{|\omega|}
(1-{\cal P}_{\rm \s})^{N_b-|\omega|} \\
B_{\rm \w}(\omega)={\cal P}_{\rm \w}^{|\eta|}(1-{\cal P}_{\rm \w})^{N_b-|\eta|}
\end{eqnarray}
and the bond occupation probabilities are
\begin{eqnarray}
{\cal P}_{\rm \s} & = & 1- \exp(-4\beta |J|) \\
{\cal P}_{\rm \w} & = & 1- \exp(-2\beta |J|).
\end{eqnarray}
The $\Delta$ and $\Gamma$ factors constrain where the two types of 
occupied bonds are allowed,
\begin{eqnarray}
\Delta(\sigma,\tau,\omega;J)&=& \left\{ \begin{array}{l}
1 \mbox{   if  for every $xy$: }  \omega_{xy}=1\rightarrow
J_{xy}\sigma_x \sigma_y>0 \mbox{ and } J_{xy}\tau_x \tau_y>0
 \\0 \mbox{ otherwise}\end{array} \right.\\
 \Gamma(\sigma,\tau,\eta)&=& \left\{ \begin{array}{l}
1 \mbox{   if  for every $xy$: }  \eta_{xy}=1\rightarrow
\sigma_x \sigma_y\tau_x \tau_y<0
 \\0 \mbox{ otherwise}\end{array} \right.
\end{eqnarray}
We refer to the $\omega$ occupied bonds as ``\s'' and the $\eta$ 
occupied bonds as ``\w''.  The $\Delta$ constraint says that \s bonds 
are allowed only if the bond is satisfied in both replicas.  The 
$\Gamma$ constraint says that \w bonds are allowed only if the bond is
satisfied in exactly one replica. 

It is straightforward to verify that the spin marginal of 
the \tr Edwards-Sokal weight is the weight for two independent Ising 
spin glasses with the same couplings,
\begin{equation}
\sum_{\{\omega\}\{\eta\}} {\cal W}(\sigma,\tau,\omega,\eta;J)=
\mbox{const} \times \exp
\left[   \beta \sum_{\{xy\}} J_{xy} (\sigma_x\sigma_y+\tau_x\tau_y)\right]
\end{equation}

\subsection{Properties of Graphical Representations for Spin Glasses}
\label{sec:properties}

Connectivity by occupied bonds in the \tr representation is related to correlations of the local spin glass order parameter,
\label{sec:pgr}
\begin{equation}
Q_x = \sigma_x\tau_x .
\end{equation}
It is straightforward to verify that 
\begin{eqnarray}
\label{eq:trcor}
\langle Q_x Q_y \rangle = 
\mbox{Prob}\{x \mbox{ and } y \mbox{ connected by even number of red bonds}\}\\
-\mbox{Prob}\{x \mbox{ and } y \mbox{ connected by odd number of red bonds}\}.\nonumber
\end{eqnarray}
As in the case of the FK representation, a minus sign complicates the
relationship between correlations and connectivity but in a conceptually
different way.  The second term in Eq.\ (\ref{eq:trcor}) is independent of
the underlying coupling in the model and is present for both spin glasses
and ferromagnetic models.

In the case of a ferromagnet, having a percolating cluster (or clusters) in
the (single replica) FK representation easily shows that there is broken
symmetry with respect to global spin flips. For example, one can impose
plus or minus boundary conditions on those boundary spins belonging to FK
percolating networks in the Edwards-Sokal joint spin-bond represenatation
and these two choices of boundary conditions give two different Gibbs
states for the spin system in the infinite volume limit.  More simply, in
the ferromagnetic case, the magnetization order parameter equals the total
density of the percolating network(s), since finite FK clusters do not
contribute.

We remark, as noted in Section~\ref{sec:intro}, that for ferromagnets (in
the absence of boundary conditions that force interfaces), the signature of
ordering is a {\it single\/} percolating cluster.  For spin glasses, the
situation is analogous, but more complicated. If there is in the CMR
graphical representation a percolating blue cluster of strictly larger
density than any other blue clusters, one can similarly show broken
symmetry. Here one can impose ``agree'' or ``disagree'' boundary conditions
between those $\sigma_x$ and $\tau_x$ boundary spins belonging to the
maximum density blue network.\footnote{We remark that this is indeed a {\it
boundary\/} condition because, in the infinite volume limit, determining
which boundary spins belong to the maximum density network does not use
information from any fixed, arbitrarily large, finite region.} In the
infinite volume limit, these two boundary conditions give different Gibbs
states for the $\sigma$-spin system (for fixed $\tau$) related to each
other by a global spin flip (of $\sigma$). However, in this case, it is not
so easy to rigorously relate (in general) the overlap $Q$ to the densities
of percolating blue networks, even if one assumes that there are exactly
two such networks with densities $D_1$ and $D_2$.   
This is because in a two-replica situation, it is not immediate that there is no
contribution from finite (non-percolating) clusters, which would be enough
to imply that $Q=D_1-D_2$. Nevertheless, this identity seems likely to be
the case, and indeed is valid for the SK model, as we discuss in the next
section of the paper.

For the TRFK representation, similar reasoning shows that the occurrence of
exactly two doubly-occupied percolating FK clusters with different
densities implies broken symmetry for the spin system~\cite{NS07} and that
$Q$ should equal (and does equal in the SK model) the density
difference.  In Sec.\ \ref{sec:results} we will present peliminary
numerical evidence that there is such a nonzero density difference below
the spin glass transition temperature for the $d=3$ EA $\pm J$ spin glass
(for both the TRFK and CMR representations).

\section{The Spin Glass on the Complete Graph}
\label{sec:cg}
The spin glass on the complete graph was introduced by Sherrington and
Kirkpatrick~(SK)~\cite{SK75}.  The $\pm J$ version of the model has
couplings given by $\pm N^{-1/2}$ where $N$ is the number of vertices on
the graph.  This scaling for the coupling strength insures that the free
energy is extensive.  In this section, we study the percolation properties
of both the Fortuin-Kastelyn and \tr representations for the SK model, In
the high temperature phase, $\beta<\beta_c=1$, both the magnetization and
the EA order parameter, $q_{EA}$, vanish.  The SK solution, valid for the high
temperature phase, yields the energy per spin, $u=-\beta/2$.  The number of
unsatisfied edges minus the number of satisfied edges is equal to
$uN^{3/2}$.  Thus, letting $f_s$ be the fraction of satisfied edges, we
have that
\begin{equation}
\label{eq:fracsat}
f_s \sim \frac{1}{2} - u N^{-1/2} \ .
\end{equation}

In the FK representation a fraction ${\cal P}_{\rm FK}= 1- \exp(-2 \beta
N^{-1/2})\approx 2 \beta N^{-1/2}$ of satisfied edges are occupied.  When
will the occupied edges first form a giant cluster and how many giant
clusters will coexist?  The theory of random graphs 
(see~\cite{Bollobas01}) can be used to answer this question.  It is
known~\cite{ER60} that a giant cluster forms in a random graph of $N$
vertices when a fraction $x/N$ of edges is occupied with $x>1$, and that
there is then a single giant cluster.  This suggests that (single replica)
FK giant clusters should form with $\beta = x N^{-1/2}$ when $x>1$, i.e.,
that the single replica FK percolation threshold is at
\begin{equation}
\label{eq:fkperc}
\beta_{{\rm FK},p}=N^{-1/2} \, .
\end{equation}
It also suggests that above this threshold, there should be a single giant
FK cluster.

Although the arguments just given are incomplete in that the satisfied
edges were treated (without justification) as though they were chosen
independently of each other, nevertheless the conclusions can be proved
rigorously as we now explain. Indeed, our rigorous analysis of the much
more interesting cases with two replicas will use very similar
arguments. The idea is to obtain upper and lower bounds for the conditional
probability that an edge $\{x_0 y_0\}$ is satisfied, given the satisfaction
status of all the other edges. If these bounds are close to each other (for
large $N$) then treating the satisfied edges as though chosen independently
can be justified.

A key point is that because of frustration, such approximate independence
is impossible if one knows too much about the signs of the couplings.
Thus, we will {\it not\/} condition on the sign of the single coupling
$J_{x_0 y_0}$ --- in fact we will consider precisely the conditional
probability of that sign given the configuration of all other couplings
$J_{xy}$ and all spins $\sigma_x$. For the $\pm J$ model that we are
considering, it is quite elementary to see first that the ratio
$Z_{+}/Z_{-}$ for the partition functions with $J_{x_0 y_0}=+N^{-1/2}$ and
$J_{x_0 y_0}=-N^{-1/2}$ satisfies
\begin{equation}
\exp(-2 \beta N^{-1/2}) \, \leq \, |Z_{+}/Z_{-}| \, \leq
\exp(2 \beta N^{-1/2}) \, ,
\end{equation}
and then that the conditional probabilities $P_{\pm}$
that $J_{x_0 y_0}=\pm N^{-1/2}$ given any configuration
of the other $J_{xy}$'s and all $\sigma_x$'s satisfy
\begin{equation}
e^{-4\beta/\sqrt{N}} \leq e^{-2\beta/\sqrt{N}}|Z_{-}/Z_{+}|
\leq P_{+} / P_{-}\leq e^{2\beta/\sqrt{N}}|Z_{-}/Z_{+}|
\leq e^{4\beta/\sqrt{N}} \, .
\end{equation}

It then follows that the conditional probabilities $P_s$ or
$P_u$ for any edge $\{x_0 y_0\}$ to be satisfied or unsatisfied
given the satisfaction status of all other edges satisfy
\begin{equation}
e^{-4\beta/\sqrt{N}} \leq P_s/P_u \leq e^{4\beta/\sqrt{N}} \, ,
\end{equation}
so that

\begin{equation}
\frac{1}{2}-{\mathrm O}(\beta/\sqrt{N}) = (e^{4\beta/\sqrt{N}}+1)^{-1}
\leq P_s \leq (e^{-4\beta/\sqrt{N}}+1)^{-1} =\frac{1}{2}+{\mathrm O} (\beta/\sqrt{N})\,.
\end{equation}
One now obtains rigorously the same conclusions as before --- i.e.,
(\ref{eq:fkperc}) is valid with a single giant FK cluster
for $\beta = \beta_N \geq x N^{-1/2}$ with any $x>1$.
Before proceeding to our detailed analysis of the situation with
two replicas, we state our main conclusions:

{\it The threshold for TRFK percolation is\/}
\begin{equation}
\beta_{{\rm TRFK},p} = 1  \ .
\end{equation}
{\it For $\beta \leq 1$, there is no giant cluster (containing a strictly
positive density, i.e., fraction of sites). For $\beta > 1$ there are
exactly two giant clusters with unequal densities. (Strictly speaking, we
do not rigorously rule out the possibility that for some choices of $\beta
> 1$, there might be only a single giant cluster, but we 
explain why that should not be so and also prove that a nonzero spin-spin
overlap rules out the possiblity of two clusters of exactly equal
density.)\/}

{\it The threshold for percolation of blue bonds in the CMR two-replica
graphical representation is\/}
\begin{equation}
\label{tworepperc}
\beta_{{\rm CMR}, p}= N^{-1/2} \, .
\end{equation}
{\it Only above that threshold are there one or more giant clusters.
The number and density of the giant clusters is determined by a second
threshold which is exactly the SK spin glass critical value $\beta_c =1$.
For $ x N^{-1/2} \leq \beta_N \leq 1$ with $x > 1$, there are
exactly two giant clusters, which have equal densities; if 
$ N^{1/2} \beta_N \to \infty$, then the two densities are both
exactly $1/2$. For $\beta_N \geq x$ with any $x>1$, there
are two giant clusters of unequal densities, whose sum is one. 
(Strictly speaking,
as in the case of TRFK percolation, we do not rigorously
rule out the possibility that for some $\beta > 1$, there might
be a single blue giant cluster, which would necessarily have density
one.)\/}

Now we explain our analysis when there are two spin replicas $\sigma$ and
$\tau$. For both TRFK percolation and for blue percolation in the CMR
graphical representation, we focus on {\it doubly satisfied\/} edges.  For
the Fortuin-Kastelyn representation, doubly satisfied edges are occupied
with probability ${\cal P}_{\rm \dfk} = [1- \exp(-2 \beta N^{-1/2})]^2 \sim
4\beta^2/N$.  For the \tr representation, doubly satisfied edges are
occupied with probability ${\cal P}_{\rm \tr} = 1- \exp(-4 \beta N^{-1/2})
\sim 4 \beta N^{-1/2}$.  The crucial new ingredient in two-replica
situations is that an edge $\{x y\}$ can be doubly satisfied only if
$\sigma_x \sigma_y \tau_x \tau_y = +1$ or equivalently if $\sigma_x \tau_x
= \sigma_y \tau_y$ (and then will be satisifed for exactly one of the two
signs of $J_{xy}$). Thus, before proceeding as in the single replica
situation, we first divide all $(\sigma,\tau)$ configurations into two
groups or sectors --- the {\it agree\/} (where $\sigma_x = \tau_x$) and the
disagree sectors (where $\sigma_x = -\tau_x$). We also denote by $N_a$ and
$N_d$ the numbers of sites in the sectors and denote by $D_a = N_a/N$ and
$D_d= N_d/N$ the sector densities (so that $D_a+D_d=1$). We note that the
spin overlap $Q$ is just
\begin{equation}
Q = \frac{1}{N} \sum_x \sigma_x \tau_x = \frac{N_a-N_d}{N} = D_a -D_d
\end{equation}
and that for $\beta \leq \beta_c =1 $, 
$Q\to 0$ as $N \to \infty$ while for $\beta > \beta_c =1 $, $Q$ is nonzero,
e.g., in the sense that $Av (<Q^2>) >0$ as $N \to \infty$,
where $Av$  denotes the average over couplings. 

We now proceed similarly to the single replica case, but separately
within the agree and disagree sectors. Letting $\bar P_\pm$ denote
the conditional probabilities that $J_{x_0 y_0} = \pm N^{-1/2}$ given
the other $J_{x y}$'s and all $\sigma_x$'s and $\tau_x$'s,
we have within either of the two sectors that
\begin{equation}
e^{-8\beta/\sqrt{N}} \leq e^{-4\beta/\sqrt{N}}|Z_{-}/Z_{+}|^2
\leq {\bar P_{+}} / {\bar P_{-}}\leq e^{4\beta/\sqrt{N}}|Z_{-}/Z_{+}|^2
\leq e^{8\beta/\sqrt{N}}
\end{equation}
so that the conditional probability {\it within a single sector\/} $P_{ds}$
for ${x_0 y_0}$ to be doubly satisfied is $(1/2) + {\mathrm O}(\beta
N^{-1/2})$. For $\beta \leq \beta_c$, we have $D_a =1/2$, $ D_d =1/2$ (in
the limit $N \to \infty$) and so in either sector, double FK percolation is
approximately a random graph model with $ N/2$ sites and bond occupation
probability $(1/2) 4 \beta^2 N^{-1} = \beta^2 (N/2)^{-1}$; thus double FK
giant clusters do not occur for $\beta^2 \leq 1$.

Blue percolation corresponds to bond occupation probability $(1/2) 4 \beta
N^{-1/2}$ $= \beta N^{1/2}(N/2)^{-1}$ and so the threshold for blue
percolation is given by~(\ref{tworepperc}).
But now there are {\it two\/} giant clusters, one in each of the two
sectors, and they are of equal density for $\beta \leq \beta_c =1$ since
$D_a =D_d$. On the other hand, for $\beta > \beta_c$, $D_a \neq D_d$ and
the two giant clusters will be of unequal density. In fact, since $\beta
N^{1/2} \to \infty$ for $\beta > \beta_c$ (indeed for any fixed $\beta>0$),
it follows from random graph theory that each giant cluster occupies the
entire sector so that $D_a$ and $D_d$ are also the cluster percolation
densities of the two giant clusters.

In the case of two-replica FK percolation for $\beta > \beta_c$, let us denote
by $D_{max}$ and $D_{min}$ the larger and smaller of $D_a$ and $D_d$, so
that $D_{max}+D_{min}=1$ and $D_{max}-D_{min}=Q$. Then for $\beta >
\beta_c$, the bond occupation probability in the larger sector is
$\beta^2(N/2)^{-1} = 2 \beta^2 D_{max} (D_{max}N)^{-1}$ with $2 \beta^2
D_{max}>1$ and there is a (single) giant cluster in that larger
sector. There will be another giant cluster (of lower density) in the
smaller sector providing $2 \beta^2 D_{min}$ ($ =\beta^2 (1-Q)$) $
>1$. Since $Q\leq q_{EA}$, for this to be the case it suffices if for
$\beta > \beta_c$,
\begin{equation}
q_{EA} < 1 - \frac{1}{\beta^2} \ .
\end{equation}
The estimated behavior of $q_{EA}$ both as $\beta \to 1+$ and as $\beta \to
\infty$~\cite{BY86} suggests that this is always valid. In any case, we
have rigorously proved that there is a unique maximal density double FK
cluster for $\beta > \beta_c$.

\section{Numerical Simulations}
\label{sec:sim}
In this section we describe numerical simulations of the Edwards-Anderson
spin glass in three dimensions to test ideas about the percolation
signature for spin glass ordering.  For comparison, we also describe
simulations of spin clusters for the $3D$ ferromagnetic Ising model.

\subsection{Methods}
We carried out simulations of the $\pm J$ Ising spin glass using a Monte
Carlo method that combines \tr cluster moves, Metropolis sweeps and
parallel tempering (replica exchange).  A similar scheme was used in a
study of the random field Ising model~\cite{MaNeCh00}.  The cluster moves
are closely related to the replica Monte Carlo algorithm introduced by
Swendsen and Wang~\cite{SwWa86,WaSw88,WaSw05} and developed
in~\cite{Houdayer01,jorg05,jorg06}. The combination of two-replica cluster moves and parallel tempering was
first introduced by Houdayer~\cite{Houdayer01}.  The new ingredient in the
present algorithm is that all of the degrees of freedom available in the
\tr representation are used.  The additional degree of freedom is
incorporated in ``grey'' moves, described below.

The parallel tempering component of the algorithm works with $R$  pairs
of replicas at equally spaced inverse temperatures. Standard temperature
exchange moves are carried out between one of the two replicas at one 
temperature and at one of the neighboring temperatures.  The \tr cluster
moves begin by identifying all singly and doubly satisfied bonds and 
occupying them with probabilities, ${\cal P}_{\rm \s} = 1- \exp(- 4 \beta )$
and ${\cal P}_{\rm \w} = 1- \exp(- 2 \beta )$, respectively.  Bonds that
are not satisfied in either replica cannot be occupied.  The occupied 
bonds determine \s and \g clusters.  Sets of sites connected by \s bonds
and singletons are considered to be \s clusters.  Sets of sites connected
by \s or \w bonds are considered to be \g clusters.  The cluster moves
proceed as follows.  For each \g cluster a random bit determines whether
to perform a \g move or not.  If a \g move is chosen, the sign of $\q$
in each \s cluster in the \g cluster is reversed.  Next, a random bit 
determines which of the two spin states to put each \s cluster in given its
$\q$ state.  For example, if $\q=1$ in a given \s cluster then, with 
equal probability, the spin state of a given site in the cluster is 
either $(++)$ or $(--)$.

The two-replica cluster component of the algorithm satisfies detailed balance 
with respect to the Edwards-Sokal weight for the \tr graphical 
representation. In addition, the two-replica cluster component of the algorithm
is, by itself, ergodic. There is a non-vanishing probability that any 
given site is a singleton cluster and is flipped to any of the four spin
states.  Thus, there is a non-vanishing probability of a transition from
any spin configuration for the pair of replicas to any other spin 
configuration in a finite number of steps.  A single pair of replicas 
will eventually approach equilibrium under two-replica cluster moves.  However,
for reasons that will become clear in Sec.\ \ref{sec:results}, the 
equilibration by two-replica cluster moves alone is very slow in three dimensions.
Thus we supplement these moves with both temperature exchange moves and
Metropolis sweeps.

The presence of very long lived metastable states makes it difficult to
gauge whether a spin glass simulation has reached equilibrium.  Here we
measure the time it takes for a spin configuration, originally at the
highest simulated temperature, to diffuse by replica exchange moves, to the
lowest temperature.  If the entire set of replicas is in equilibrium and if
the replica pair at the highest temperature is rapidly equilibrated, then
this first passage time estimates the time it takes to obtain an independent
sample at the lowest temperature. We assume that the mean first passage
time is comparable to the equilibration time for the full set of replicas
though it is conceivable that equilibration time is much longer than the
mean first passage time.  For the 12$^3$ system, the largest system studied
here, the mean first passage time is of the order of one hundred MC sweeps
and does not vary greatly from one realization to another so we believe
that the system is well equilibrated.

The two-replica cluster moves complicate the ability to keep track of a single spin
configuration as it diffuses in temperature space. If there are two giant
clusters and, in one of the replicas, both or neither of the giant clusters
are flipped then the identity of the spin configuration is unaffected 
(though it may have suffered an overall spin flip). On the other hand,
if one giant cluster is flipped and the other not then the spin 
configurations of the replicas are swapped.  Below the \tr percolation
transition, this same rule is applied to the two largest clusters but 
the identity of the spin configuration is effectively lost in one two-replica move.

\subsection{Improved Estimators}
One advantage of cluster algorithms in data collection is the existence of
improved estimators~\cite{Wolff} in the graphical representation.  For
example, in the Swendsen-Wang algorithm, one can obtain the magnetization
and magnetic susceptibility from the cluster configurations.  The
magnetization is the average size of the largest cluster and the
susceptibility is proportional to the sum of the squares of the cluster
sizes. Since each occupied bond configuration corresponds to many spin
configurations, the variance of observables measured from the bond
configurations is less than for the same observables measured in the spin
configuration leading to smaller error bars for the same amout of
computational work.  Improved estimators exist within the \tr
representation for the spin glass order parameter and susceptibility. For
example, the order parameter $Q$ can be obtained from the percolating grey
cluster if it exists and is unique.  In particular, the local order
parameter $Q_x$ summed over the sites of the percolating grey cluster
should equal the $Q$ of the whole system since the contributions of small
grey clusters vanishes after averaging over the possible spin states of
these clusters.

\subsection{Simulation results}
\label{sec:results}

We simulated the three-dimensional $\pm J$ Edwards-Anderson model on skew
periodic cubic lattices for system sizes $6^3$, $8^3$, $10^3$ and $12^3$.
For each size we simulated 20 inverse temperatures equally spaced in the
range $\beta=0.16$ to $\beta=0.92$.  Since there are two replicas for each
temperature, the total number of replicas simulated was 40.  A recent
estimate of the phase transition temperature of the system is
$\beta_c=0.89\pm 0.03$ \cite{KaKoYo06}.  For each size we simulated 100
realizations of disorder for 50,000 Monte Carlo sweeps of which the first
$1/4$ of the sweeps were for equilibration and the remaining $3/4$ for data
collection.  One Monte Carlo sweep consists of a cluster move for the pair
of replicas at every temperature, a Metropolis sweep for each replica and a
temperature exchange attempt for each temperature.  The quantities that we
measure are the fraction of sites in the largest \s cluster, ${\cal C}_1$
and second largest \s cluster, ${\cal C}_2$ and the number of \s \tr
wrapping cluster, $w_{\rm \tr}$, and the number of \dfk ``wrapping''
clusters, $w_{\rm \dfk}$.  A cluster is said to wrap if it is connected
around the system in any of the three directions.

\begin{figure}
 \includegraphics{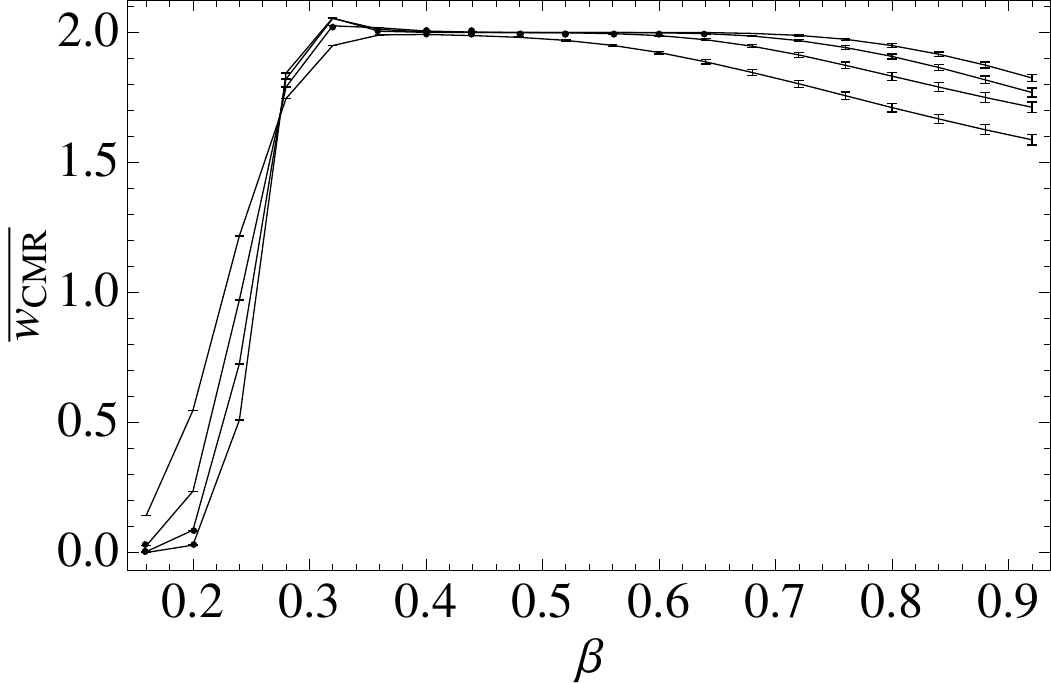}
\caption{Average number of wrapping \tr clusters, $\overline{w_{\rm CMR}}$ vs.\  $\beta$ for the 3D EA model.}
\label{fig:wrap}       \end{figure}

\begin{figure}
 \includegraphics{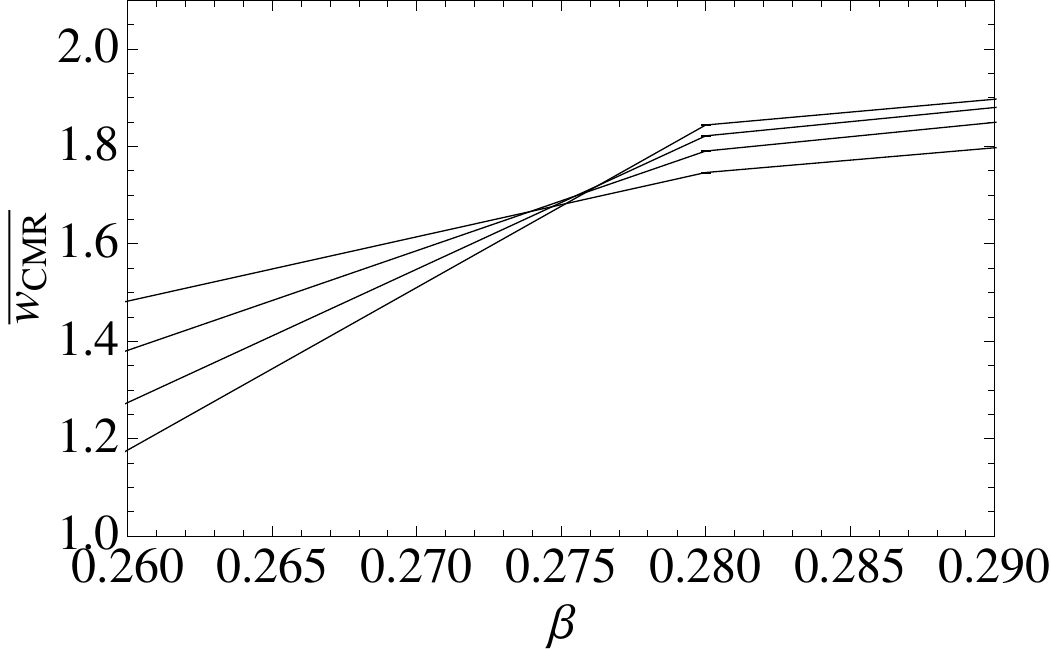}
\caption{Same as Fig.\ \ref{fig:wrap}, magnified near the \tr percolation transition}
\label{fig:wrapclose}      
\end{figure}

Figure \ref{fig:wrap} shows the average number $\overline{w_{\rm CMR}}$ of
\tr blue wrapping clusters 
as a function of inverse
temperature $\beta$.  The curves are ordered by system size with largest
size on the bottom for the small $\beta$ and on top for large $\beta$.  The
data suggests that there is a percolation transition at some $\beta_{{\rm
CMR},p}$.  For $\beta> \beta_{{\rm CMR},p}$ there are {\em two} wrapping
clusters while for $\beta<\beta_{{\rm CMR},p}$ there are none. Near and
above the spin glass transition at $\beta_c \approx 0.89$ the expected
number of wrapping clusters falls off but the fall-off diminishes as system
size increases.  This figure suggests that in the large size limit there
are exactly two spanning clusters near the spin glass transition both above
and below the transition temperature.  Figure \ref{fig:wrapclose} is a
magnification of Fig.\ \ref{fig:wrap} near the \tr percolation transition.
The crossing points identify the percolation transition as $\beta_{{\rm
CMR},p} \approx 0.275$.  This value is close to the FK percolation
transition for the $\pm J$ EA model $\beta_{{\rm FK},p} \approx 0.26$
reported in~\cite{ArCoPe91}.  A more careful study would be needed to test
the hypothesis that $\beta_{{\rm CMR},p}>\beta_{{\rm FK},p}$.

\begin{figure}
 \includegraphics{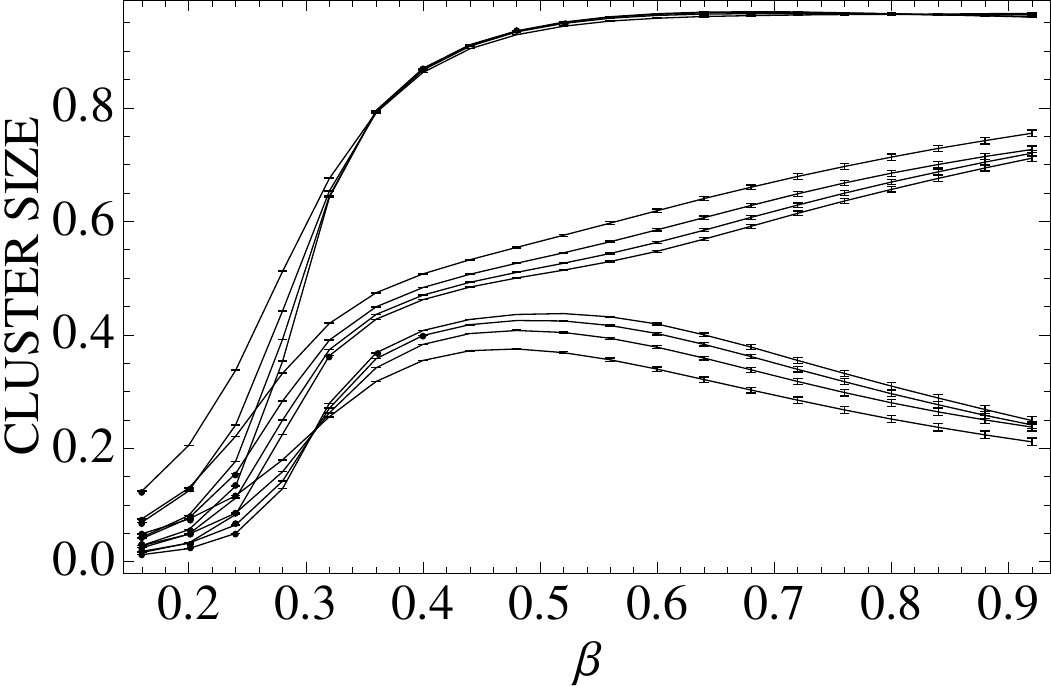}
\caption{${\cal C}_1$ (middle set), ${\cal C}_2$ (bottom set) and ${\cal C}_1+{\cal C}_2$ (top set) vs.\
$\beta$ for the  \tr graphical representation for the 3D EA model.}
\label{fig:clustersize}       
\end{figure}

Figure \ref{fig:clustersize} shows the fraction of sites in the largest \tr
blue cluster, ${\cal C}_1$, second largest \tr blue cluster, ${\cal C}_2$
and the sum of the two, ${\cal C}_1+{\cal C}_2$. The middle set of four
curves is ${\cal C}_1$ for sizes $6^3$, $8^3$, $10^3$ and $12^3$, ordered
from top to the bottom at $\beta=0.5$.  The bottom set of curves is ${\cal
C}_2$ with systems sizes ordered from smallest on bottom to largest on top
at $\beta=0.5$.  The difference between the fraction of sites in the two
largest clusters, ${\cal C}_1-{\cal C}_2$ is approximately the spin glass
order parameter.  As the system size increases, this difference decreases
below the transition suggesting that ${\cal C}_1={\cal C}_2$ for
$\beta<\beta_c$ in the thermodynamic limit.  On the other hand, the sum of
the two largest clusters is quite constant independent of system size. Near
the transition, approximately 96\% of the sites are in the two largest
clusters.

The large fraction of sites in the two largest clusters makes the \tr
cluster moves inefficient.  If all sites were in the two largest clusters
then the cluster moves would serve only to flip all spins in one or both
clusters or exchange the identity of the two replicas.  Equilibration
depends on the small fraction of spins that are not part of the two largest
clusters.  One of the reasons that bond diluted spin glasses are more
efficiently simulated using two-replica cluster algorithms is the smaller
fraction of sites in the two largest clusters.  We have carried out
simulations on the same bond diluted Ising spin glass studied by
J\"{o}rg~\cite{jorg05,jorg06}. This model has 55\% of the couplings set to
zero and 45\% set to $\pm 1$.  Near the phase transition, we find that only
87\% of the sites are contained in the two largest clusters instead of the
96\% found in the undiluted spin glass.

\begin{figure}
\includegraphics{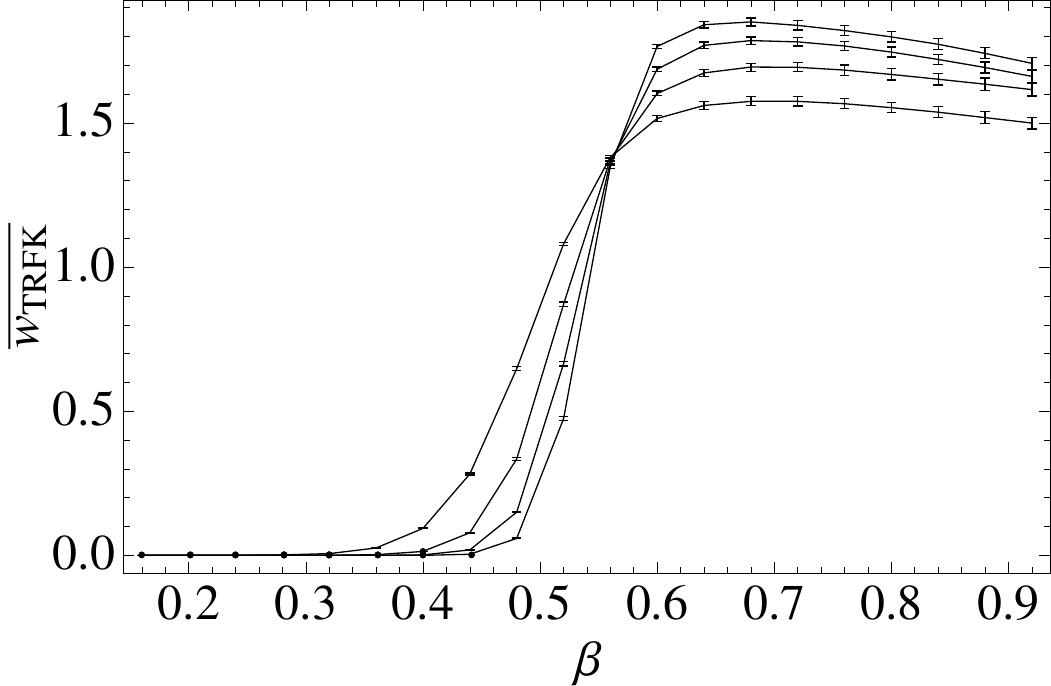}
\caption{Average number of doubly occupied wrapping Fortuin-Kastelyn
clusters, $\overline{w_{\rm \dfk}}$ vs.\ $\beta$ for the 3D EA model.}
\label{fig:fkwrap}       
\end{figure}

\begin{figure}
 \includegraphics{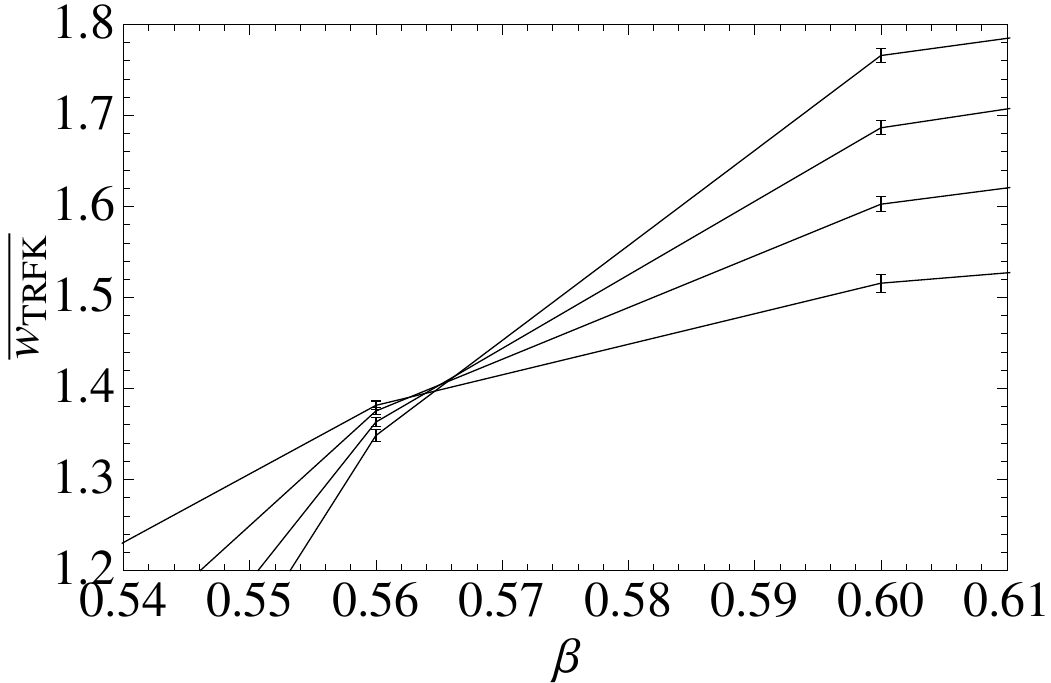}
\caption{Same as Fig.\ \ref{fig:fkwrap}, magnified near the \dfk percolation transition}
\label{fig:fkwrapclose}       
\end{figure}

Figure \ref{fig:fkwrap} show the average number of wrapping \dfk clusters 
$\overline{w_{\rm \dfk}}$
as a function of inverse temperature.  The largest system size is on the
bottom for the small $\beta$ and on top for the large $\beta$.  As for the
case of \tr clusters, the data suggests a transition at some $\beta_{{\rm
\dfk},p}$ from zero to two wrapping \dfk clusters.  Although the number of
\dfk wrapping clusters is significantly less than two for all $\beta$ and
all system sizes, the trend in system size suggests that it might approach
two for large systems and $\beta>\beta_{{\rm \dfk},p}$.  Figure
\ref{fig:fkwrapclose} shows a close up of the transitions region and the
crossing points give the inverse percolation temperature as $\beta_{{\rm
\dfk},p} \approx 0.565$

\begin{figure}
\includegraphics{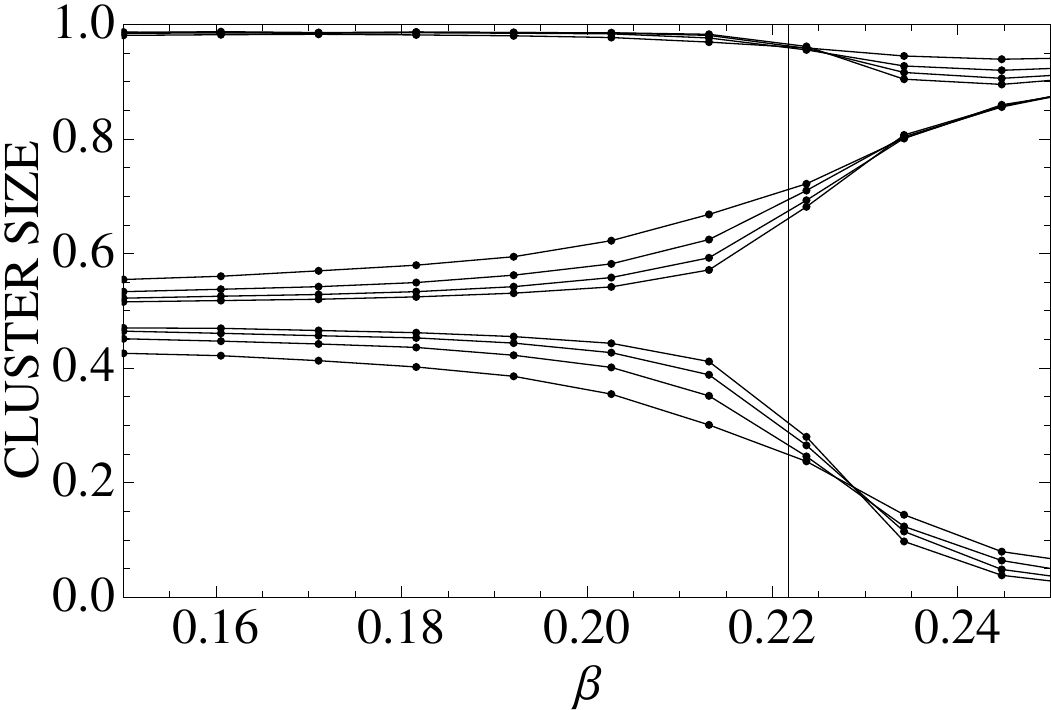}
\caption{The size of the largest cluster of satisfied bonds (middle set), the second largest
cluster (bottom set) and the sum of the two largest clusters (top set) vs.\ $\beta$ for the three-dimensional ferromagnetic Ising model.}
\label{fig:ferroclustersize}       
\end{figure}

The percolation signature for both \tr and \dfk clusters is qualitatively
similar in three dimensions.  In both cases two giant clusters with
opposite values of the local order parameter appear at a temperature
substantially above the phase transition temperature.  In the high
temperature phase, the two giant clusters have the same density and the
phase transition is marked by the onset of different densities of the two
clusters.  This scenario is not unlike what is expected in the
ferromagnetic Ising model for the graph defined by satisfied bonds.  On the
cubic lattice in the high temperature phase we expect two giant clusters of
up and down spins of equal density.  The ferromagnetic Ising phase
transition is marked by the onset of different densities of the two
clusters.  Figure \ref{fig:ferroclustersize} shows the largest and second
largest cluster of satisfied bonds and the sum of the two for the
ferromagnetic Ising model. The system sizes are the same as for the spin
glass simulations: $6^3$, $8^3$, $10^3$ and $12^3$. The vertical axis is
located at the critical temperature.  This figure is qualitatively similar
to the results for the two largest clusters in the \tr representations. In
both cases, for these system sizes, the phase transition is quite rounded
in the sense that a difference in density develops well before the
transition and the transition itself cannot be identified by looking at the
size of the clusters.  The difference in the density of the two clusters is
thus not a sharp indicator of the phase transition for small system sizes.

\section{Discussion}

In this paper we have proposed a new percolation-theoretic approach towards
understanding the nature of the spin glass phase transition.  It is based
on the Fortuin-Kasteleyn~\cite{KF69,FK72} random cluster method (and some
variants), which has been enormously useful in analyzing phase transitions
in ferromagnets, but had much less impact on systems such as spin glasses
until now (see, however,~\cite{WaSw88,jorg05,jorg06}).

There are a number of advantages to our approach.  First, and most
obviously, it sheds new light on the nature of the spin glass phase
transition, particularly its geometric aspects.  For example, it provides
at least a qualitative explanation of why the EA spin glass on a simple
planar lattice doesn't have broken spin flip symmetry at positive
temperature (see below).  Second, it indicates a possible new framework
towards an eventual rigorous proof of an EA spin glass phase transition (at
least in sufficiently high dimensions), while providing a basis for
numerical work to explore, and hopefully resolve, the issue of the lower
critical dimension.  Finally, it helps to emphasize fundamental differences
between phase transitions in spin glasses as opposed to more conventional
systems such as ferromagnets.

Two different representations, each involving two replicas, were used to
apply an FK type formalism --- the
Chayes-Machta-Redner(CMR)~\cite{CMR98,CRM98} representation and the two-replica FK (TRFK) representation previously considered by
Newman-Stein~\cite{NS07}.  It is likely that other representations can also
be used, as long as they involve the overlap of independent replicas.
While various details will differ, as described in the text, the essential
--- and somewhat surprising --- feature appears to be that the spin glass
transition coincides with the emergence of percolating clusters of unequal
densities.

Numerical results for the $d=3$ EA spin glass seem to indicate that this
occurs, as $T$ is lowered below $T_c$, as the breaking of symmetry in the
equal densities of both doubly occupied TRFK and blue CMR clusters that
already percolate {\it above\/} the transition temperature.  For the SK
model, on the other hand, what occurs just above $T_c$ is
representation-dependent: in the TRFK representation, there is no doubly
occupied percolation at all above $T_c$, while in the CMR representation
there are two equal-density blue clusters just above $T_c$, similar to the
situation in the $d=3$ EA model.  This difference in behavior above $T_c$
may arise from the peculiarities of the SK model, and a similar
representation-dependence may not occur in short-range models.

Finally, we speculate about the nature of the lower critical dimension.
Our numerical results are consistent with prior
studies~\cite{BaCrFe00,KaKoYo06} indicating the appearance of broken
spin-flip symmetry in the EA model in three dimensions.  If the percolation
signature scenario proposed here is correct for short-range models, it
would help explain why there is no spin glass transition leading to broken
spin-flip symmetry on simple planar lattices: two dimensions does not
generally provide enough ``room'' for two disjoint infinite clusters to
percolate.  However, a system that is infinite in extent in two dimensions
but finite in the third might be able to support two percolating clusters,
with unequal densities at low temperature.  This and other possibilities
will be explored in future work.

\begin{acknowledgments}
We thank the Aspen Center for Physics where some of this work was
performed.  JM was supported in part by NSF DMR-0242402.  He thanks New
York University and Alan Sokal for hosting his sabbatical during which time
this work was begun.  CMN was supported in part by NSF DMS-0102587 and
DMS-0604869.  DLS was supported in part by NSF DMS-0102541 and DMS-0604869.
Simulations were performed on the Courant Institute of Mathematical
Sciences computer cluster.
\end{acknowledgments}

\bibliographystyle{unsrt}

\bibliography{refs61707}

\end{document}